\documentclass[prb,aps,amsmath,amssymb,showpacs,reprint]{revtex4-2}
\usepackage{amssymb,graphicx,float,dcolumn,bm,subfigure,xcolor}
\usepackage[titletoc,title]{appendix}
\usepackage[dotinlabels]{titletoc}
\usepackage{subfigure}
\usepackage{braket}
\usepackage{nicefrac}
\usepackage[normalem]{ulem}

\usepackage[
pdfpagelabels,%
colorlinks=true,breaklinks=true,%
linkcolor=purple,urlcolor=blue,citecolor=blue,%
]{hyperref}

\newcommand{\be}{\begin{equation}}
\newcommand{\ee}{\end{equation}}

\newcommand{\ba}{\begin{eqnarray}}
\newcommand{\ea}{\end{eqnarray}}
\renewcommand{\vec}[1]{{\textbf{\textit{#1}}}}

\usepackage{amsmath}

\begin{document}

\title{Crystalline Solutions of Kohn-Sham Equations in the Fractional Quantum Hall Regime}
\author{Yayun Hu}
\thanks{These two authors contributed equally.}
\author{Yang Ge}
\thanks{These two authors contributed equally.}
\author{Jian-Xiao Zhang}
\author{J. K. Jain}
\affiliation{Physics Department, 104 Davey Laboratory, Pennsylvania State University, University Park, PA 16802}

\begin{abstract}
A Kohn-Sham density functional approach has recently been developed for the fractional quantum Hall effect, which maps the strongly interacting electrons into a system of weakly interacting composite fermions subject to an exchange correlation potential as well as a density dependent gauge field that mimics the ``flux quanta" bound to composite fermions. To get a feel for the role of various terms, we study the behavior of the self-consistent solution as a function of the strength of the exchange correlation potential, which is varied through an {\it ad hoc} multiplicative factor. We find that a crystal phase is stabilized when the exchange correlation interaction is sufficiently strong relative to the composite-fermion cyclotron energy. Various properties of this crystal are examined.
\end{abstract}

\maketitle

\section{Introduction}\label{Background}

The Kohn-Sham (KS) density functional theory (DFT) treats an interacting electronic system by mapping it into a system of non-interacting electrons moving in an effective single-particle KS potential~\cite{Giuliani08}. The validity of the KS scheme relies on the assumption of the form of the universal Hohenberg-Kohn energy functional, or the exchange-correlation (XC) functional, the search for which has motivated extensive studies. Despite the success of DFT, for example at the level of local density approximation (LDA), its application to strongly interacting systems has been more challenging and requires more sophisticated treatments~\cite{Seidl99b,Gori-Giorgi09,Malet12,Yu2016}. 

\begin{figure}[t]
\includegraphics[width=\columnwidth]{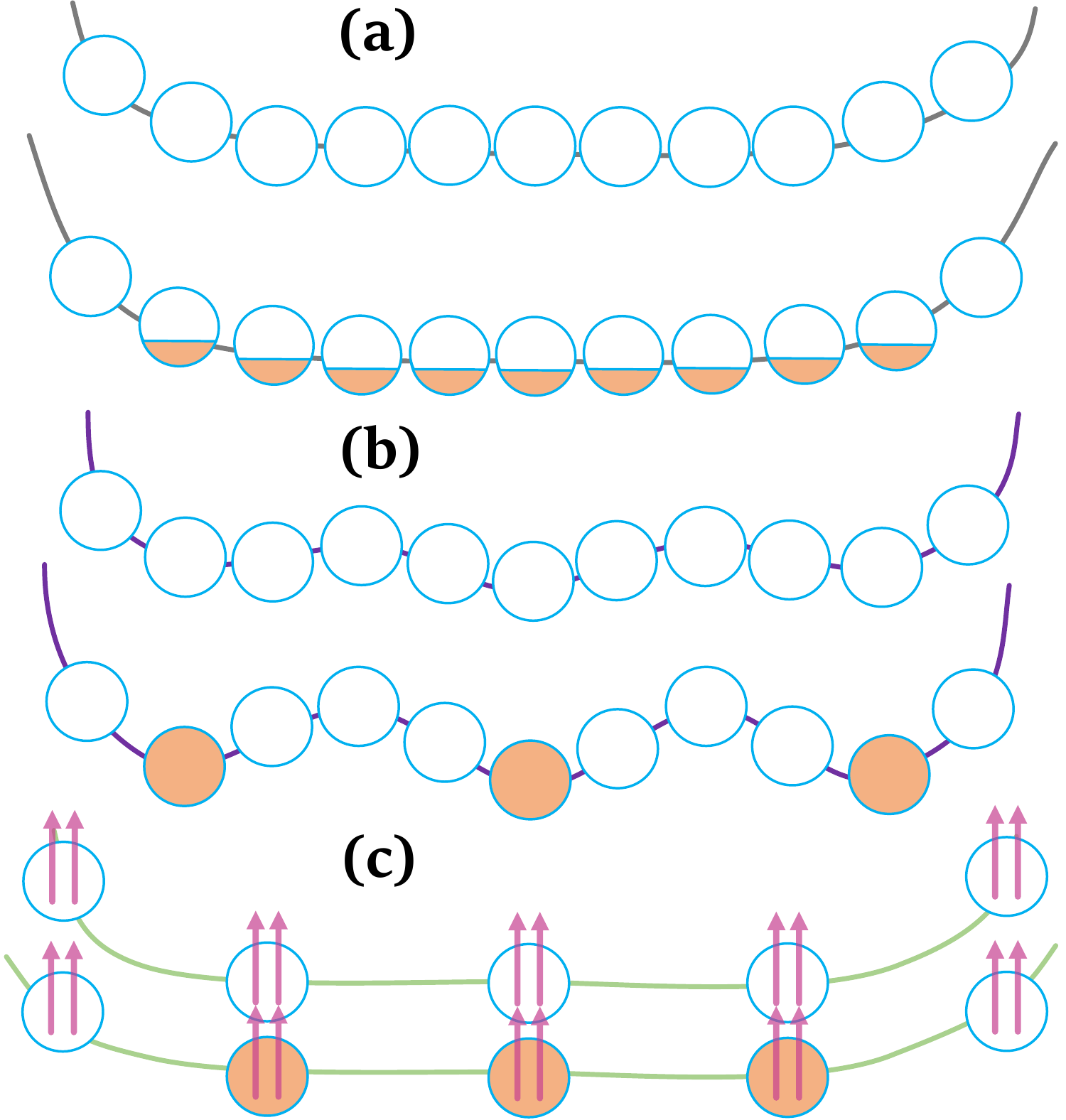}
\caption{Schematic illustration of our KS approach for the FQHE. (a) The real system under consideration, taken to be at $1/3$-filling in an external confining potential $V_{\rm ext}$. In the bulk, each orbital is occupied, on average, by 1/3 of an electron. The fractional occupation is illustrated by the partial coloring of the otherwise empty circles that represent unoccupied orbitals. 
(b) The KS spectrum of an auxiliary problem of non-interacting electrons in a KS potential $V_{\rm KS}$, with the renormalized LLs illustrated by purple lines. The orbitals are either fully occupied or empty. (c) The occupation configuration of the auxiliary system of emergent composite fermions (shown as electrons plus two flux quanta) in the composite-fermion Landau level (called $\Lambda$ level) spectrum. The integer quantum Hall effect of CFs in a KS$^*$ potential $V_{\rm KS}^*$ can reproduce the uniform density of the real system in (a). The DFT treatment in this paper maps the real system of FQHE to an auxiliary system in (c). }\label{FQHEmap}
\end{figure}

The system of interest to us is the fractional quantum Hall effect (FQHE)~\cite{Tsui82}, which occurs when electrons in two dimensions are subjected to a strong perpendicular magnetic field which quenches their kinetic energy and as a result enhances the effects of Coulomb interaction. 
Very few papers~\cite{Ferconi95,Heinonen95,Zhang14c,Zhang15b,Zhao17,Hu19} have been written applying DFT to the FQHE since its discovery over the past four decades. 
The difficulty of applying DFT to FQHE traces back to the construction of KS DFT by mapping into non-interacting electrons, because the non-interacting system possesses a large degeneracy. That can be seen by considering the canonical Hamiltonian that describes the bulk of the FQHE sample:
\begin{equation}
\mathcal{H}_{\rm LLL}={V}_{\rm ee}=\sum_{i<j}^{N}\frac{e^2}{\epsilon |\vec{r}_i-\vec{r}_j|}\;. \label{bareHam}
\end{equation}
Here $N$ electrons have been taken to be confined in the lowest Landau level (LLL), as appropriate in the limit of very large magnetic fields. The quantity $\epsilon$ is the dielectric constant of the host material. In Eq.~\eqref{bareHam}, the external potential due to a neutralizing background has been suppressed, and the constant kinetic energy of the LLL has been dropped. In the absence of interaction, the ground state for free electrons has a large degeneracy that counts all the possible ways of occupying $N$ of the $N_{\phi}$ Landau level (LL) orbitals, where $N_{\phi}$ is the single-particle degeneracy of the LLL. The switching-on of the interaction leads to correlated ground states at certain special values of the filling factor $\nu=N/N_{\phi}$. This correlated ground states are certain very complicated linear superpositions of the large number of basis functions, such that the average occupation of each single particle orbital is fractional (equal to $\nu=N/N_{\phi}$). 
In the KS DFT treatment of the FQHE, mapping the problem into free electrons in a KS potential is equivalent to replacing ${V}_{\rm ee}$ by ${V}_{\rm KS}$ in the Hamiltonian of Eq.~\eqref{bareHam}. The main point is that the KS solution picks out a single Slater determinant, which is insufficient to describe the FQHE state. In fact, the KS solution can only be a nonuniform state that locally describes an integer quantum Hall effect, rather than a state in which each LL orbital has a fractional occupation. This physics has been illustrated in Fig.~\ref{FQHEmap}. In the DFT literature, two different directions have been developed to treat strong correlation effects: the improvement of XC and the replacement of the Slater determinant in the Kohn-Sham system by a multiconfiguration function~\cite{Yu2016}. In the FQHE, one can imagine addressing this issue with the help of an exchange correlation potential that has cusps at certain densities~\cite{Ferconi95,Heinonen95}; Ref.~\onlinecite{Ferconi95} implements an ensemble average over successive iterations to produce, on average, fractional occupation of LL orbitals. 

We circumvent this problem by mapping interacting electrons into an auxiliary system of non-interacting composite fermions~\cite{Hu19}, referred to as the KS* system below whenever it is necessary to differentiate it from the standard KS system of non-interacting electrons. Composite fermions (CFs)~\cite{Jain89,Jain07,Halperin20}, often thought of as the bound state of electrons and an even number (2p) of flux quanta [see Fig.~\ref{FQHEmap}(c)], are the emergent weakly-interacting particles of the FQHE. In particular, the FQHE of electrons is an integer quantum Hall effect (IQHE) of CFs. The advantage of this method is that the integrally occupied orbitals of composite fermions represent fractionally occupied levels of electrons. 
The KS* equation for composite fermions has been derived in a way analogous to the standard formulation in the KS scheme of DFT.

It is important in the CF DFT to incorporate properly the non-local gauge interaction between CFs, which arises due to the attached fluxes. The long-range nature of the gauge interaction is crucial for capturing the topological properties of the FQHE. An advantage of CF DFT is that it simplifies the modeling of the XC energy. Like in any DFT method, the exact form of the XC enengy of CFs is not known and must be approximated. However, since CFs are weakly interacting, it is reasonable to assume that their XC energy is a smooth function of density within an LDA. (This corresponds to an XC energy for electrons that has cusps at the Jain fillings~\cite{Ferconi95,Heinonen95,Price96}.) Applications of the CF DFT scheme have obtained not only the ground state density and energy, but also the topological properties of the excitations~\cite{Hu19}, including their fractional charge and fractional braiding statistics, which are robust against the specific choice of the XC potential~\cite{Hu21}.

The goal of the present work is to investigate the behavior of the solution as a function of the strength of the XC potential.  The XC potential of composite fermions should be dependent on the quantum well width and LL mixing, but the dependence is likely to be complex, and we have not studied that here. Instead, we vary the strength of the XC potential through an {\it ad hoc} multiplicative factor.  To allow for most general solutions we implement the CF DFT in a manner that does not impose any symmetry on the solutions. (In Ref.~\onlinecite{Hu19}, we had applied CF DFT to a rotationally symmetric system, while also assuming, for the convenience of numerical calculation, that the solution also has a rotational symmetry.) Our primary finding is that the liquid state of composite fermions, which occurs for weak XC potentials, yields to a crystal phase when the magnitude of exchange-correlation potential is raised relative to the CF cyclotron energy.

One may ask to what extent our study applies to realistic systems. It is possible that a given choice of the XC potential may correspond to some interaction between electrons, but we have not made any attempt to identify the latter. Our results in this work are only to be viewed at a qualitative level.
Previous studies have considered, in a variational calculation, transitions between liquid and crystal states of composite fermions as a function of the filling factor or Landau level mixing~\cite{Archer13,Zhao18}. It is plausible that increasing the strength of the XC potential enhances mixing between the Landau levels of composite fermions, called $\Lambda$ levels ($\Lambda$L), and thus causes a crystal in the same fashion as LL mixing has been shown to do~\cite{Zhao18}. 
There is an important difference, however. Those studies compared energies of liquid and crystal states of composite fermions carrying different numbers of vortices. In our current study, we assume that the state is always described in terms of a given species of composite fermions (which will be assumed to be composite fermions carrying two quantized vortices below). In other words, the liquid as well as the crystal states in our study are states of the same species of composite fermions.  It is therefore unclear how our results relate to previous studies or to experiments.

The plan of the manuscript is as follows. In Sec.~\ref{HowToCalculate} we review the Kohn-Sham equations. In Sec.~\ref{SquareLattice} we explain in detail how to numerically solve the Kohn-Sham equations using a finite difference method on a square lattice. As an application of our method, in Sec.~\ref{results}, we study the ground state density as a function of the strength of the exchange-correlation energy. The effects of temperature, weak disorder, the form of XC energy, and system size are also considered. We summarize our findings in Sec.~\ref{summary}.

\section{KS$^*$ equations for composite fermions}\label{HowToCalculate}

We consider the following Hamiltonian of a FQHE system in a 2D $x$-$y$ plane with an external potential $V_{\rm ext}$:
\be
\mathcal{H}={T}_{\rm ee}+{V}_{\rm ee}+{V}_{\rm ext}\;,\label{fullHam}
\ee
where ${T}_{\rm ee}=\sum_{i=1}^{N}\frac{1}{2m_b}(\vec{p}+\frac{e}{c}\vec{A})^2 $ is the kinetic energy operator, $m_b$ is the band mass of electrons and $\vec{A}$ is the vector potential due to a uniform external magnetic field $\vec{B}=B \vec{e}_{z}=\nabla\times \vec{A}(\vec{r})$ along the $z$ direction.

Following the magnetic-field DFT (BDFT)~\cite{Grayce94,kohn04,Tellgren12}, the total energy functional $E_{\rm tot}$ of the FQHE system can be expressed as a functional of the ground state density $\rho(\vec{r})$ as: 
\be
E_{\mathrm{tot}}[\rho]=E_{\rm K}[\rho]+E_{\rm xc}[\rho]+E_{\rm H}[\rho]+\int d^2\vec{r} V_{\rm ext}(\vec{r})\rho(\vec{r})\;.\label{Interpretation}
\ee
Here the Hartree energy $E_{\rm H}[\rho]$ takes the standard form:
\begin{equation}\label{EH}
E_{\rm H}(\rho)=\frac{1}{2}\int d^2\vec{r} d^2\vec{r}'\frac{\rho(\vec{r})\rho(\vec{r}')}{\epsilon|\vec{r}-\vec{r}'|}\;.
\end{equation}
In order to study the FQHE in the LLL, throughout this paper we have defined the non-interacting kinetic energy functional to be $E_{\rm K}[\rho]\equiv \frac{1}{2}\hbar\omega_B \int d^2 \vec{r}\rho(\vec{r})=\frac{N}{2}\hbar\omega_B$, where $\omega_B=\frac{eB}{m_bc}$ is the cyclotron frequency. The electron XC energy functional $E_{\rm xc}$ is defined by Eq.~\eqref{Interpretation}. It can be equivalently defined through a constrained search formalism~\cite{Levy79,Lieb83}
\be
E_{\rm xc}[\rho]\equiv\min_{\Psi\rightarrow \rho(\vec{r})}\langle \Psi|{T}_{\rm ee}+{V}_{\rm ee}|\Psi\rangle-E_{\rm K}[\rho]-E_{\rm H}[\rho]\;,\label{ExcDef}
\ee
which is further simplified by using $\langle \Psi_{\rm LLL}|{T}_{\rm ee}|\Psi_{\rm LLL}\rangle=\frac{N}{2}\hbar\omega_B$ in the LLL:
\be
E_{\rm xc}[\rho]=\min_{\Psi_{\rm LLL}\rightarrow \rho(\vec{r})}\langle \Psi_{\rm LLL}|{V}_{\rm ee}|\Psi_{\rm LLL}\rangle-E_{\rm H}[\rho]\;,\label{ExcDefLLL}
\ee
where the many-body wave function $\Psi_{\rm LLL}$ searches for an energy minimum of ${V}_{\rm ee}$ within the LLL Hilbert space. We adopt Eq.~\eqref{ExcDefLLL} in the following.

The above formulation is the standard version of the Hohenberg-Kohm (HK) theorem. 
Here $E_{\rm xc}$ depends on the external magnetic field $\vec{B}$ in the BDFT formalism, but is otherwise a universal functional of density that does not depend on the external potential.

We next construct an auxiliary KS* system of CFs (rather than electrons), which is not typical of the standard KS scheme, but is within the formulation of the generalized KS scheme~\cite{Seidl96} and the concepts in the standard KS scheme apply as usual. 
We imagine that there exists a reference system that consists of non-interacting CFs, whose ground state density is the same as the ground state density of the FQHE system and is expressed as the sum of the contribution from the occupied KS orbitals as
\be
\rho(\vec{r})=\sum_{\alpha} c_{\alpha}|\psi_{\alpha}(\vec{r})|^2\;,\label{rhodef}
\ee
where $c_{\alpha}$ is the occupation number of the KS orbital labeled by $\alpha$. We then define the ``non-interacting" kinetic energy $T_{\rm s}^*[\rho]$ of CFs as
\be\label{KSstartotalEKrequireHKform}
T_{\rm s}^*[\rho]=\sum_{\alpha}\langle\psi_\alpha | T^* | \psi_\alpha\rangle,
\ee
where the kinetic energy operator $T^*$ of CFs is
\be
T^*= \frac{1}{2m^*}\left(\vec{p}+\frac{e}{c}\vec{A}^*(\vec{r};[\rho])\right)^2\;.
\ee
The important physics of CFs is incorporated through the density-dependent effective vector potential $\vec{A}^*(\vec{r})$, or effective magnetic field $\vec{B}^*$, through 
\be
\nabla \times \vec{A}^*(\vec{r})=B^*(\vec{r})\vec{e}_{z}=\left[ B- 2\rho(\vec{r})\phi_0\right]\vec{e}_{z}\;.\label{BstarDef1}
\ee
As in the standard KS scheme, a further connection between the FQHE system and the KS system of CFs is made by rewriting $E_{\rm xc}$ as
\be
E_{\rm xc}[\rho]=T_{\rm s}^*[\rho]+E^*_{\rm xc}[\rho]\;,\label{ExcPartition}
\ee
which also defines $E^*_{\rm xc}[\rho]$ as the XC energy of CFs. To clarify, despite the absence of the kinetic energy of electrons, the kinetic energy of CFs arises from Coulomb interaction and is included as part of $E_{\rm xc}[\rho]$. In Ref.~\onlinecite{Hu19}, the CF XC energy was approximated in LDA as:
\be
E_{\rm xc}^*(\rho)=\varsigma\int d^2\vec{r}\left[a\nu^{1/2}+(b-f/2)\nu+g\right]\rho(\vec{r})\;,\label{EXCStarintroduction0602}
\ee
with parameters
$a=-0.78$, $b=0.28$, $f=0.33$, $g=-0.050$ in units of $\frac{e^2}{\epsilon l_B}$, and $\nu(\vec{r})=2\pi l^2_B\rho(\vec{r})$ is the local filling factor, where $l_B=\sqrt{\frac{\hbar c}{eB}}$ is the magnetic length. The parameter $\varsigma$ will be used to control the strength of the XC potential; $\varsigma=1$ corresponds to the choice in Ref.~\onlinecite{Hu19}. The first term $a\nu^{1/2}$ in $E_{\rm xc}^*$ is chosen to match with the known classical value of energy of the Wigner crystal in the limit $\nu\rightarrow 0$~\cite{Bonsall77}, and the coefficients for higher orders of $\nu$ are chosen to fit the electronic XC energies that are obtained using trial wave functions at the Jain fillings $\nu=n/(2n+1)$. This XC form is suited for the filling factor range $1/3<\nu<1/2$, but we will use it uncritically for arbitrary filling factors below. The value of $g$ gives a constant energy offset and does not affect the KS orbitals or the ground state densities. (We note the a similar multiplicative factor to tune the strength of the XC potential has been used in other contexts, for example for a model hydrogen molecule~\cite{Holst2019}.)

Minimization of $E_{\rm tot}$ is achieved by variations with respect to the KS orbitals, which leads to the KS* equation:
\begin{widetext}
\be
H^*\psi_{\alpha}(\vec{r})=\left[T^*
+V_{\rm H}(\vec{r})+V_{\rm ext}(\vec{r})+V_{\rm xc}^{*}(\vec{r})+V^*_{\rm T}(\vec{r}) \right] \psi_{\alpha}(\vec{r}) = \epsilon_{\alpha} \psi_{\alpha}(\vec{r})\;,
\label{singleCFKSintroduction}
\ee
\end{widetext}
where the Hartree potential takes the standard form
\begin{equation}
V_{\rm H}(\vec{r})=\frac{e^2}{\epsilon}\int d^2\vec{r}'\,\frac{\rho(\vec{r}')}{|\vec{r}-\vec{r}'|}\;.\label{VH}
\end{equation}
The CF XC potential is obtained through $V_{\rm xc}^*(\vec{r})\equiv\delta E^*_{\rm xc}/\delta \rho(\vec{r})$ as:
\be
V_{\rm xc}^*(\vec{r})=\varsigma\left[\frac{3}{2}a\nu^{1/2}(\vec{r})+(2b-f)\nu(\vec{r})+g\right].\label{VXCStarintroduction}
\ee
In the KS potential experienced by the CFs, there is a non-standard term $V^*_{\rm T}$ that is defined as:
\begin{equation}
V^*_{\rm T}(\vec{r})=\sum_{\alpha}c_{\alpha}\langle\psi_{\alpha}|\frac{\delta {T}^*}{\delta \rho(\vec{r})}|\psi_{\alpha}\rangle\;,\label{DeltaTStarDefintroduction}
\end{equation}
which comes from the density-dependence of the vector potential $\vec{A}^*$ inside the kinetic energy operator ${T}^*$ of CFs. This term is typically much smaller than the other terms in the KS potential. In particular, the effect of $V^*_{\rm T}$ is irrelevant for the topological properties~\cite{Hu19,Hu21}. It is worth emphasizing that while the kinetic energy operator of electrons, ${T}_{\rm ee}$, is absent in the KS* equation, the kinetic energy of CFs enters the KS* equation, and incorporates the non-perturbative effect of the Coulomb interaction.

At finite temperatures, $\{c_{\alpha}\}$ and the chemical potential $\mu$ are determined by 
\begin{eqnarray}
c_{\alpha} &=& \frac{1}{1+\exp[(\epsilon_{\alpha}-\mu)/k_\mathrm{B}\tau]}\;, \label{FDdistribution1}\\
N &=& \sum_{\alpha} c_{\alpha}\;.\label{FDdistribution2}
\end{eqnarray}
The occupation number $c_{\alpha}$ reduces to either $0$ or $1$ in the limit of zero temperature.

\section{Numerical procedure for solving the Kohn-Sham equations}\label{SquareLattice}

In this section, we outline the numerical procedure adopted to find the KS solutions. We show how the finite-difference method is implemented on a discretized lattice. We also discuss the algorithm applied in successive iterations to achieve convergence. 

\subsection{Choice for the magnetic vector potential $\vec{A}^* $}

We consider a rectangular 2D system of sides $L_x$ and $L_y$. 
We discretize the system into a lattice and label each lattice point as $ \vec{r}=(x,y)$, where $x=i a_x$, $y=j a_y$, with $i=1,2,\cdots,N_x$ and $j=1,2,\cdots,N_y$ and the lattice constants are $a_x=L_x/N_x$ and $a_y=L_y/N_y$, respectively. This allows a discretization of the physical quantities. 

The effective magnetic field for CFs is given by
\begin{equation}
\vec{B}^*(\vec{r})=B^*(\vec{r})\vec{e}_{z}=\left[1-2\nu(\vec{r}) \right]B\vec{e}_{z}\;, \label{BstarDef2}
\end{equation}
which is equivalent to Eq.~\eqref{BstarDef1}.
In order to write down the KS Hamiltonian explicitly, we pick the symmetric gauge for the bound flux of CFs. The vector potential $\vec{A}^*(\vec{r})$ reads
\begin{eqnarray}\label{AstarComponentSum}
	\vec{A}^*(\vec{r}) & = & \int d^2\vec{r}' \, \frac{ B^*(\vec{r}')} {2 \pi | \vec{r}-\vec{r}' |^2 } \, {\vec{e}_z \times (\vec{r}-\vec{r}')} \nonumber\\
	& = & \sum_{\vec{r}^\prime {\ne \vec{r}}}\frac{B^*(\vec{r}^\prime)a_xa_y}{2\pi |\vec{r}-\vec{r}^\prime|^2} \, (y^\prime-y,x-x^\prime).
\end{eqnarray}
The sum over $\vec{r}^\prime$ extends over all space. This choice satisfies the Coulomb gauge condition $\nabla\cdot \vec{A}^*=0$, which can be checked explicitly and implies the commutation relation $[\vec{p}, \vec{A}^*]=0$. 

\subsection{The discretized Hamiltonian}

The discretized form of the KS Hamiltonian is straightforward. We show here the form explicitly for the $V_{\mathrm{T}}$ term, which is non-standard and also the most complex. 
We proceed as follows
\begin{eqnarray}
&&V_{\rm T}(\vec{r})=\sum_{\alpha}c_{\alpha}\langle\psi_{\alpha}|\frac{1}{2m^*}\frac{\delta \left(\vec{p}'+\frac{e}{c}\vec{A}^*(\vec{r}')\right)^2}{\delta \rho(\vec{r})}|\psi_{\alpha}\rangle\label{GeneralDeltaTStarDef1}\\
&&=\sum_{\alpha}c_{\alpha}\langle\psi_{\alpha}|\frac{e}{m^* c}\frac{\delta \vec{A}^*(\vec{r}')}{\delta \rho(\vec{r})}\cdot\left(\vec{p}'+\frac{e}{c}\vec{A}^*(\vec{r}')\right)|\psi_{\alpha}\rangle\label{GeneralDeltaTStarDef4}\\
&&=\frac{\hbar eB}{2m^*c}\sum_{\alpha}c_{\alpha}\int d^2 \bar{\vec{r}}'\,\psi^*_{\alpha}(\bar{\vec{r}}')\nonumber \\ 
&&\left[\frac{\bar{y}-\bar{y}'}{|\bar{\vec{r}}'-\bar{\vec{r}}|^2}\left(-i\frac{\partial}{\partial\bar{x}'}+\bar{A}_x^*(\bar{\vec{r}}')\right)\right. \nonumber \\
&&\left.
 +\frac{\bar{x}'-\bar{x}}{|\bar{\vec{r}}'-\bar{\vec{r}}|^2}\left(-i\frac{\partial}{\partial\bar{y}'}+\bar{A}_y^*(\bar{\vec{r}}') \right)\right]\psi_{\alpha}(\bar{\vec{r}}')\;,
\end{eqnarray}
where $\langle\psi_{\alpha}|{O}(\vec{r}')|\psi_{\alpha}\rangle \equiv \int d^2 \vec{r}'\,\psi^*_{\alpha}(\vec{r}') {O}(\vec{r}')\psi_{\alpha}(\vec{r}')$, $\bar{\vec{r}}=\vec{r}/l_B$, and $\bar{\vec{A}}^* = e \vec{A}^* l_B/c$.
In Eq.~\eqref{GeneralDeltaTStarDef4}, $\left(\vec{p}'\cdot\frac{\delta \vec{A}^*(\vec{r}')}{\delta \rho(\vec{r})}\right)=0$ is used, which can be checked explicitly by noticing that
\begin{equation}
\frac{\partial}{\partial x'}\frac{\delta A_x^*(\vec{r}')}{\delta \rho(\vec{r})}=-\frac{\partial}{\partial y'}\frac{\delta A_y^*(\vec{r}')}{\delta \rho(\vec{r})}=\frac{2\phi_0}{\pi }\frac{(y-y')(x'-x)}{|\vec{r}'-\vec{r}|^4}\;,\label{Coulomb1}
\end{equation}
where we have used 
\begin{equation}
\delta [B^*(\vec{r}')]/\delta \rho(\vec{r})= -2\phi_0\delta(\vec{r}-\vec{r}')\;,
\end{equation}
and
\begin{equation}
\frac{\delta \vec{A}^*(\vec{r}')}{\delta \rho(\vec{r})}= \frac{\phi_0}{\pi |\vec{r}'-\vec{r}|^2}(y'-y,x-x')\;.
\end{equation}
The conversion $\frac{\hbar eB}{2m^*c}=0.0010\frac{e^2}{\alpha_{m^*}\epsilon l_B}$ is used in our numerical calculation, where $\alpha_{m^*}$ relates the CF mass to electron mass $m_e$ by $m^*=\alpha_{m^*}\sqrt{B[T]}m_e$. We take $\alpha_{m^*}=0.08$, which is a good approximation for theoretical transport gaps~\cite{Jain07}.

The Hartree potential is calculated using a discretized form of Eq.~\eqref{VH}. To avoid the singularity in the self energy at $\vec{r}=\vec{r}'$, we replace point charge by a uniformly distributed charge on a square region of size $a_x \times a_y$ centered around $\vec{r}$. 

In this paper, we consider an external potential $V_{\mathrm{ext}}$ generated by a uniform positive background charge inside a circular region around the origin. For a system of $N$ electrons, the background charge density is chosen as $\rho_{\mathrm{b}}=\frac{\nu_{\mathrm{b}}}{2\pi l^2_B}$ with a radius $R_{\mathrm{b}}=\sqrt{\frac{2N}{\nu_{\mathrm{b}}}}l_B$, where $\nu_{\mathrm{b}}$ is the average ion filling factor $\nu_{\mathrm{b}}=2\pi l^2_B \rho_{\mathrm{b}}$. We make sure that the rectangle $L_x\times L_y$ is chosen to be large enough so as to comfortably enclose the electron system.

\subsection{Numerical procedure for iterations}\label{iteration}

We obtain the self-consistent solution of Eq.~\eqref{singleCFKSintroduction} using the following iterative procedure. (i) We start with an input density $\rho_{\rm in}$. (ii) We obtain $T^*$ and $V^*_{\rm KS}(\vec{r})$ on the left-hand side of Eq.~\eqref{singleCFKSintroduction}, diagonalize the Hamiltonian to obtain the KS$^*$ orbitals, and determine the output density $\rho_{\rm out}=\sum_{\alpha} c_{\alpha}|\psi_{\alpha}|^2$ according to Eq.~\eqref{rhodef}, \eqref{FDdistribution1} and \eqref{FDdistribution2}. Note that we work with a fixed particle number, and therefore need to adjust the chemical potential suitably in each iteration. (iii) The relative difference $\frac{\Delta N}{N}$ between the input and output $\rho$, where $\Delta N=\int |\rho_{\rm in}-\rho_{\rm out}|d^2 \vec{r}$, is called the absolute difference. We accept $\rho_{\rm out}$ as converged if the relative differences between any of the two output densities for $2000$ successive iterations satisfy  $\frac{\Delta N}{N}<0.001$. This ensures that the solution is stable and not altered by further iterations. We find that the energy also converges when the above criterion is satisfied.
(iv) If $\rho$ has not converged, we prepare new input density $\rho_{\rm in}$ by mixing some output density into the previous input: $\rho_{\rm in}\rightarrow \eta\rho_{\rm in}+(1-\eta)\rho_{\rm out}$, where the mixing coefficient is $\eta\geq 0.9$. The choice of $\eta$ close to one helps avoid the so-called {\it occupation sloshing}, which can occur due to the large degeneracy in our system~\cite{Woods2019}.  We iterate the process until convergence is reached. It is worth mentioning that the calculation of $V_{\rm T}$ requires the information of the KS orbitals. $V_{\rm T}$ is set to zero in the initial input, but in later iterations, it is necessary to also mix the input and output $V_{\rm T}$ in the same way as the mixing of density in each iteration in order to ensure convergence. We always start at a sufficiently high temperature, where convergence is straightforward, and slowly go to lower temperatures, while using the converged density of the previous temperature as the input.

The KS Hamiltonian needs to be updated in each iteration. We notice that a direct calculation of the $\vec{A}^*$, $V_{\rm T}$ and $V_{\rm H}$ terms on each lattice site using for-loops can be time consuming. To increase efficiency, we have utilized the convolution algorithm that is available in the Intel\textsuperscript{\textregistered} Math Kernel Library (MKL) to calculate these terms. For the diagonalization of the Hamiltonian, we use the Feast algorithm~\cite{Polizzi09}, which is also available in the MKL and can take a guess of eigenstates as input to increase efficiency. This is suitable for our purpose because the KS orbitals from the previous iteration serve as a good guess of eigenvectors for the new diagonalization. In this paper, the typical system we consider has $N=40$, $L_x=L_y=35$, $N_x=N_y=210$. The corresponding Hamiltonian is a sparse matrix with a dimension of 44100. With $\eta= 0.95$, the convergence takes several thousand to tens of thousands of iterations depending on whether the converged density is liquid-like or crystal-like. Liquid-like solutions are largely uniform in the bulk and converge quickly. In contrast, crystal-like solutions require significantly larger number of iterations for convergence, in order to adjust the position and shape of the crystalline sites. The corresponding computation time can range from half a day for liquid to one week for crystal solutions respectively for the above typical system size in a single cluster node with 10 cores. 

\begin{figure*}[t]
\includegraphics[width=6.5in]{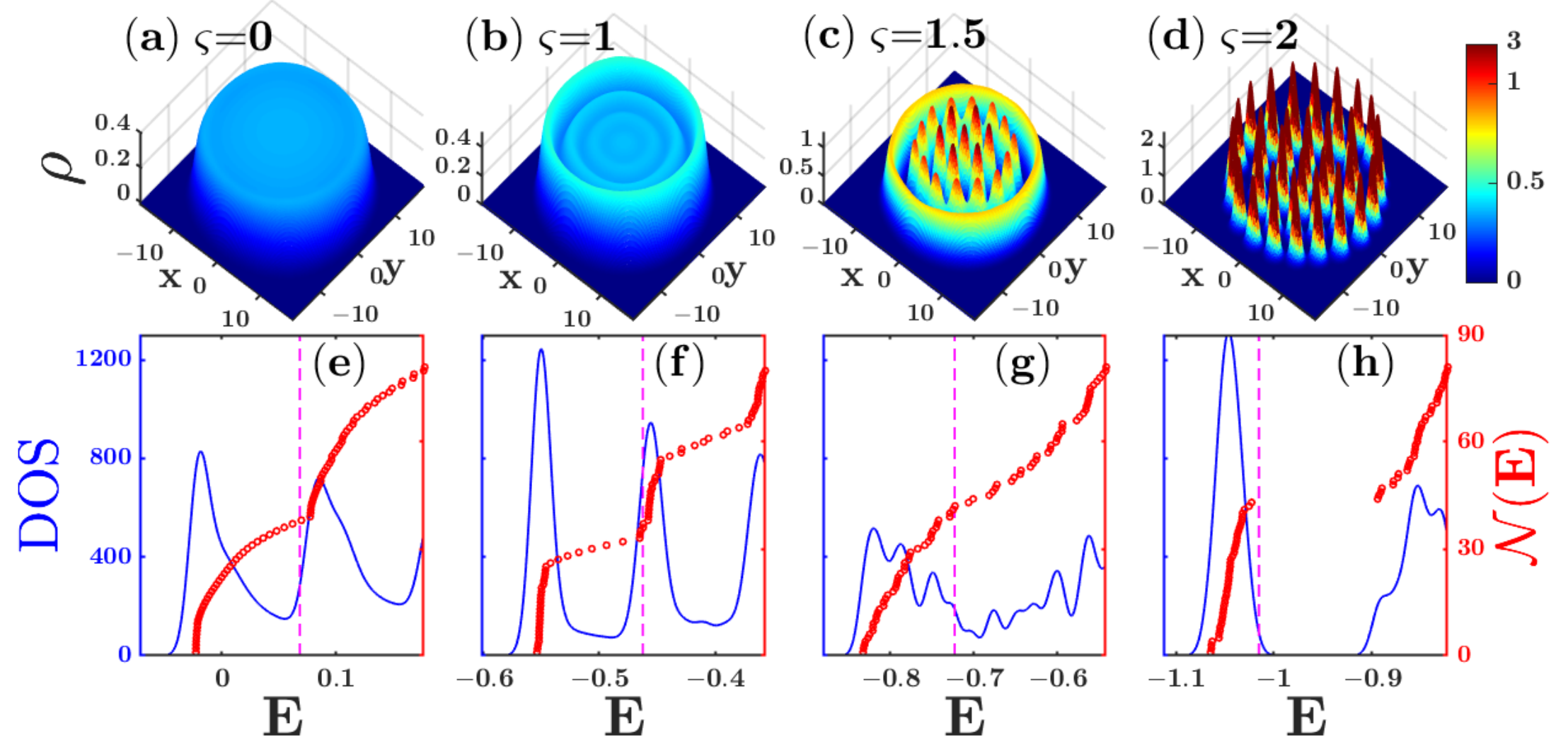}
\caption{The ground state density ($\rho$) and the density of states (DOS) of KS* orbitals for a system of $N=40$ electrons with average filling $\nu=1/3$ at temperature $k_{\rm B}\tau=0.01\frac{e^2}{\epsilon l_B}$. The system size is $L_x=L_y=35\; l_B$, and $N_x=N_y=210$. Panels (a-d) depict how the density varies as a function of the exchange-correlation (XC) potential, whose strength is tuned by the prefactor $\varsigma$; these panels correspond to $\varsigma=0, 1, 1.5, 2$. The density is quoted in units of $(2\pi l_B^2)^{-1}$. The FQHE liquid evolves into a crystal with increasing XC energy. Panels (e-f) show the corresponding density of states, $\rho_E$. The cumulative state count $\mathcal{N}(E)$ gives the number of Kohn-Sham orbitals below energy $E$. Only the lowest 85 orbitals are shown. The dashed line marks the location of the Fermi energy. All the energies are in units of $\frac{e^2}{\epsilon l_B}$. }\label{XCcompare}
\end{figure*}

\section{Results}\label{results}

The validity of our DFT results depends on the accuracy of the choice of the XC energy for CFs. We restrict the approximation of $E_{\rm xc}^*$ to the level of LDA and the form of XC energy in Eq.~\eqref{EXCStarintroduction0602} is obtained by fitting to the ground state energies of the uniform systems in the filling factor range $1/3<\nu<1/2$, which is the range where composite fermions carrying two flux quanta are relevant. Since the CFs are weakly interacting, it is reasonable to use a smooth fitting curve for the XC energy, although there has been no study of the exact constraints~\cite{Pittalis11,Dufty11} on the proper choice of the CF XC energy. Eq.~\eqref{EXCStarintroduction0602} is not unique and a different form has been applied in Ref.~\onlinecite{Zhao17}. These slightly different fitting forms do not influence the ground state energy as well as the topological properties when the system is in the filling factor range $1/3<\nu<1/2$. However, we will use this form for the XC energy uncritically for all filling factors, and all of our results are subject to this approximation.

An important point for our discussion below is that the form of the exact $E_{\rm xc}^*$ also depends on various physical factors~\cite{Price96}. For example, one possible factor is LL mixing, which is absent in the theoretical limit of very strong magnetic fields but is relevant for typical magnetic fields and can be quite significant. With LL mixing, one can reformulate the problem in terms of electrons still residing in the LLL but with an effective interaction, which is less repulsive than the Coulomb interaction at short distances. A similar correction arises due to finite width of the quantum well.  In principle, one then needs to evaluate the CF cyclotron energy and CF XC energy for the effective interaction, which is likely to change the relative importance of the two terms. We have not made a realistic determination of these effects. 

We will tune the strength of the XC energy $E_{\rm xc}^*$ in Eq.~\eqref{EXCStarintroduction0602} by varying $\varsigma$. Notice that $E_{\rm xc}^*$ is negative, so the XC potential increases in magnitude when $\varsigma>1$. We refer to $\varsigma\rightarrow \infty$ and $\varsigma\rightarrow 0$ as the strong and weak XC energy limits, respectively.

\subsection{Appearance of a crystal phase}

We consider a problem with rotational symmetry by choosing the external potential to be generated by a uniform positive charge in a circular region around the origin, as explained in Sec.~\ref{iteration}. The ground state densities for a system of $N=40$ are shown in Fig.~\ref{XCcompare}(a-d) for certain choices of $\varsigma$. The corresponding density of states (DOS) are shown in Fig.~\ref{XCcompare}(e-h). Here we assume a small temperature of $k_B\tau=0.01\frac{e^2}{\epsilon l_B}$, which is less than $10\%$ of the cyclotron gap of $\Lambda$Ls and is useful for finding converged solutions. (Temperature dependence of KS solutions is discussed later.) In the weak XC energy limit ($\varsigma=0$), the ground state density of electrons almost perfectly screens the background density, due to the dominance of the Hartree term and the external potential. The spectrum of the KS solutions shows the formation of $\Lambda$Ls, where the positions of the lowest two $\Lambda$Ls can be seen from the two peaks in the DOS plot in Fig.~\ref{XCcompare}(e).

In order to obtain a smooth curve for DOS, we have replaced the $\delta(E-\epsilon_i)$ in the standard definition of DOS, $\rho_{E}=\sum_i \delta(E-\epsilon_i)$ (where $\epsilon_i$ is the eigenvalue of KS orbitals sorted by $\epsilon_1<\epsilon_2<\ldots$, increasing with $i$), by a Gaussian to define:
\be
\rho_{E}=\sum_i\exp[-(E-\epsilon_i)^2/\sigma^2]/\sqrt{2\pi}\sigma\;,
\ee
where $\sigma=0.008\frac{e^2}{\epsilon l_B}$ throughout this paper. The discrete points of $\{\epsilon_i\}$ can be seen from the plot of the cumulative state count
\be
\mathcal{N}(E)=\int_{-\infty}^{E}\sum_i \delta(E'-\epsilon_i)dE'\;.
\ee

When $\varsigma=1$ in Eq.~\eqref{EXCStarintroduction0602}, the total density shows stronger oscillations near the edge but still respects a rotational symmetry. In particular, the density profile near the edge of the system first shoots up before it comes down to zero, which is also seen in results from exact diagonalization (ED) in a rotationally symmetric system~\cite{Tsiper01}. For the stronger XC potential of $\varsigma=1.5$, the bulk becomes a crystal; the absence of a crystal at the boundary is a finite temperature effect, as discussed below. The system fully crystalizes when $\varsigma=2$. The formation of crystalline structures breaks the rotational symmetry of the system, which is allowed in our numerical method where no symmetry is assumed. (In contrast, the calculations in Ref.~\onlinecite{Hu19} choose the angular momentum as a good quantum number and reduce the 2D system to effectively a 1D system along the radial direction; the results therein are rotationally symmetric by construction regardless of the strength of the XC potential or the choice of the CF mass. Rotation symmetry is also imposed in the ED calculation in the disk geometry~\cite{Tsiper01}.) 
The results are stable and driven by the XC energy. We discuss in Appendix~\ref{error} that our results, in particular the appearance of a crystal phase, are not a numerical artifact of discretization and lattice configuration.

\subsection{Nature of the crystal phase}

The competition between the correlated Wigner crystal state and the liquid state has been studied theoretically in many articles~\cite{Lam84,Levesque84,Price93,Platzman93,Filinov01,Jeon04a,Jeon04b,Zhao18}. In particular, Ref.~\onlinecite{Zhao18} studies the role of LL mixing and finds that the FQHE liquid yields to a crystal when the LL mixing is large. However, there is an important difference between the crystal found in that study and that in our study. In Ref.~\onlinecite{Zhao18}, the $n/(2n+1)$ FQHE liquid of composite fermions carrying two vortices freezes into an {\it electron} crystal with increasing LL mixing. In our study, on the other hand, we obtain a CF crystal of composite fermions carrying two vortices~\cite{Jeon04a,Jeon04b}. 

The Wigner crystallization of a system of 2D electrons in a circularly symmetric confining potential has been studied in Ref.~\onlinecite{Filinov01}. They find that crystallization occurs in two stages: first in the radial ordering, and then in the angular ordering. The situation is similar to our findings, though we have CF crystals rather than electron crystals. We calculate the evolution of variances in bulk density along the radial and the azimuthal directions respectively as we increase $\varsigma$ (results not shown). For a small $\varsigma$ ($\varsigma<1$), both variances remain negligible in the liquid phase. When $\varsigma$ is larger than a threshold value, the radial variance first increases significantly; beyond a greater threshold, the azimuthal variance also increases abruptly. This indicates a two-stage crystallization in our results. 

It is interesting to ask if our crystal is an example of the so-called Hall crystal~\cite{Kivelson86,Halperin86,Fradkin99}, which is the quantum Hall effect counterpart of the putative supersolid phase of $^4$He atoms. Because our crystal is a correlated crystal of composite fermions, it may appear to be a promising candidate for the Hall crystal phase. One feature that may distinguish the Hall crystal from the Wigner crystal is that, in the former, the number of particles per unit cell is not necessarily an integer~\cite{Tevsanovic89}. We find, for all cases we have studied, that the number of crystal sites in our KS solution at the smallest temperature is equal to the number of composite fermions (which also justifies the term crystal rather than a charge density wave). Another character of the Hall crystal is that, similarly to the Hall liquid state, it hosts chiral edge states. We find that for our crystal, there are no gapless edge states; this is indicated by the presence of a gap at the chemical potential.  We thus conclude that our crystal phase is generically not a Hall crystal, but we do not rule out the possibility that the Hall crystal state could be stabilized for some forms of the XC energy.

\begin{figure*}[t]
\includegraphics[width=6.5in]{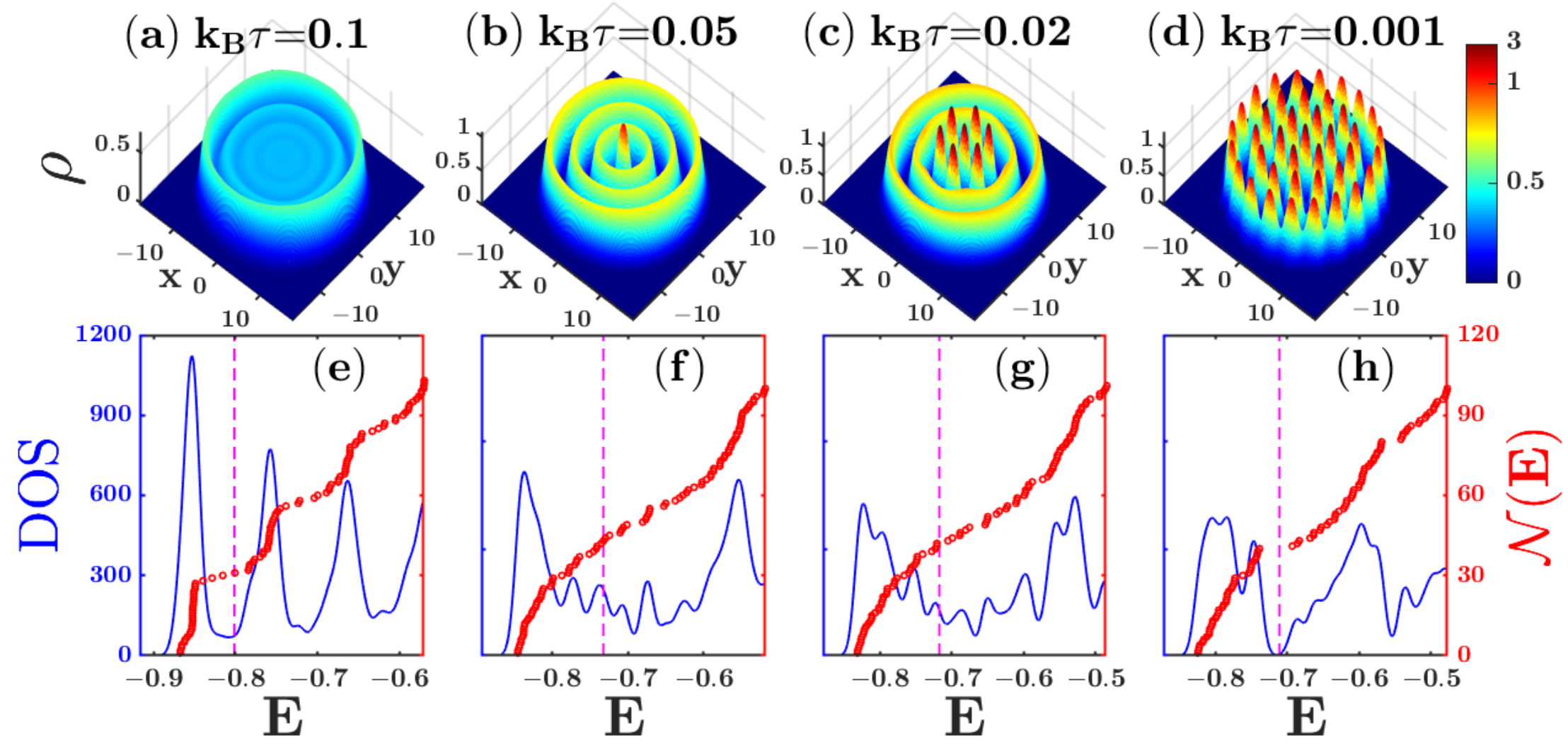}
\caption{The temperature dependence of the ground state density ($\rho$) and the density of states (DOS) of KS* orbitals for a system with $N=40$ particles at average filling $\nu=1/3$. We choose the XC potential with $\varsigma=1.5$. The density is quoted in units of $(2\pi l_B^2)^{-1}$. The panels (a-d) correspond to temperatures $k_{\rm B}\tau=0.1, 0.05, 0.02, 0.001$ in units of $\frac{e^2}{\epsilon l_B}$. The panels (e-f) show the corresponding density of states and the cumulative state count for Kohn-Sham solutions in (a-d). Only the lowest 100 orbitals are shown. Other parameters are the same as those in Fig.~\ref{XCcompare}.}\label{kBTcompare}
\end{figure*}

\subsection{Effects of temperature, disorder, form of XC interaction and system size}

We ask how temperature influences the density in our calculation. Fig.~\ref{kBTcompare} depicts the evolution of a system with $\varsigma=1.5$ as a function of temperature. At the lowest temperature of $k_B\tau=0.001\frac{e^2}{\epsilon l_B}$, the system is crystalline (panel d). As the temperature is raised, the system melts from the edge into the bulk and becomes a liquid-like state when $k_B{\tau}=0.1\frac{e^2}{\epsilon l_B}$, which is approximately the value of the $\Lambda$L gap ($\approx 0.11\frac{e^2}{\epsilon l_B}$), as can be seen from the DOS plot in Fig.~\ref{kBTcompare}(e). It has been proposed that the re-entrant of a solid state can occur in certain parameter regimes when the temperature of a liquid state is raised~\cite{Platzman93}; we have not explored that physics in our calculations. 

Next, we test the stability of the results against a weak disorder. We consider onsite disorder by adding to the external potential a term $\sum_i\delta V_{\rm ext}(\vec{r}_i)$, where $V_{\rm ext}(\vec{r}_i)$ is randomly chosen according to a uniform distribution in the range $[-W,W]$, where $W$ is the strength of disorder. In the $\nu=1/3$ state, we find that both the liquid and the crystal states largely remain unaffected for a disorder strength of up to $W = 0.01 \frac{e^2}{\epsilon l_B}$, although we expect that the phase boundary will be slightly modified by disorder~\cite{Price93}.

One may ask how the detailed form of the XC energy/potential influences the results. For that we consider another form of the XC energy $E_{\rm xc}^{*\prime}=\varsigma\int d^2\vec{r}[-0.61\nu^{0.39}-0.165\nu]\rho(\vec{r})$, which gives an XC potential $V_{\rm xc}^{*\prime}=\varsigma(-0.85\nu^{0.39}-0.33\nu)$. (These forms are different approximations for the exact energies in the range $1/3<\nu<1/2$, but have significant differences outside this range.) The qualitative behavior, namely a liquid for small $\varsigma$ and a crystal at large $\varsigma$ is also seen for the new XC potential. However, the phase boundaries are different; for example, the low-temperature KS solution is a crystal for $V_{\rm xc}^{*\prime}$ with $\varsigma=1$. 
(We note that both forms of XC energy produce a uniform liquid state in the bulk for $\varsigma=1$ when the system is constrained to be rotationally symmetric~\cite{Hu19}.) 

We have also studied the effects of the CF mass $m^*$, and the density rings and crystal sites emerge for a large $m^*$ that decreases the $\Lambda$L gap. This suggests that the formation of $\Lambda$Ls is also important for the stability of the liquid phase.

\begin{figure}[t]
\includegraphics[width=\columnwidth]{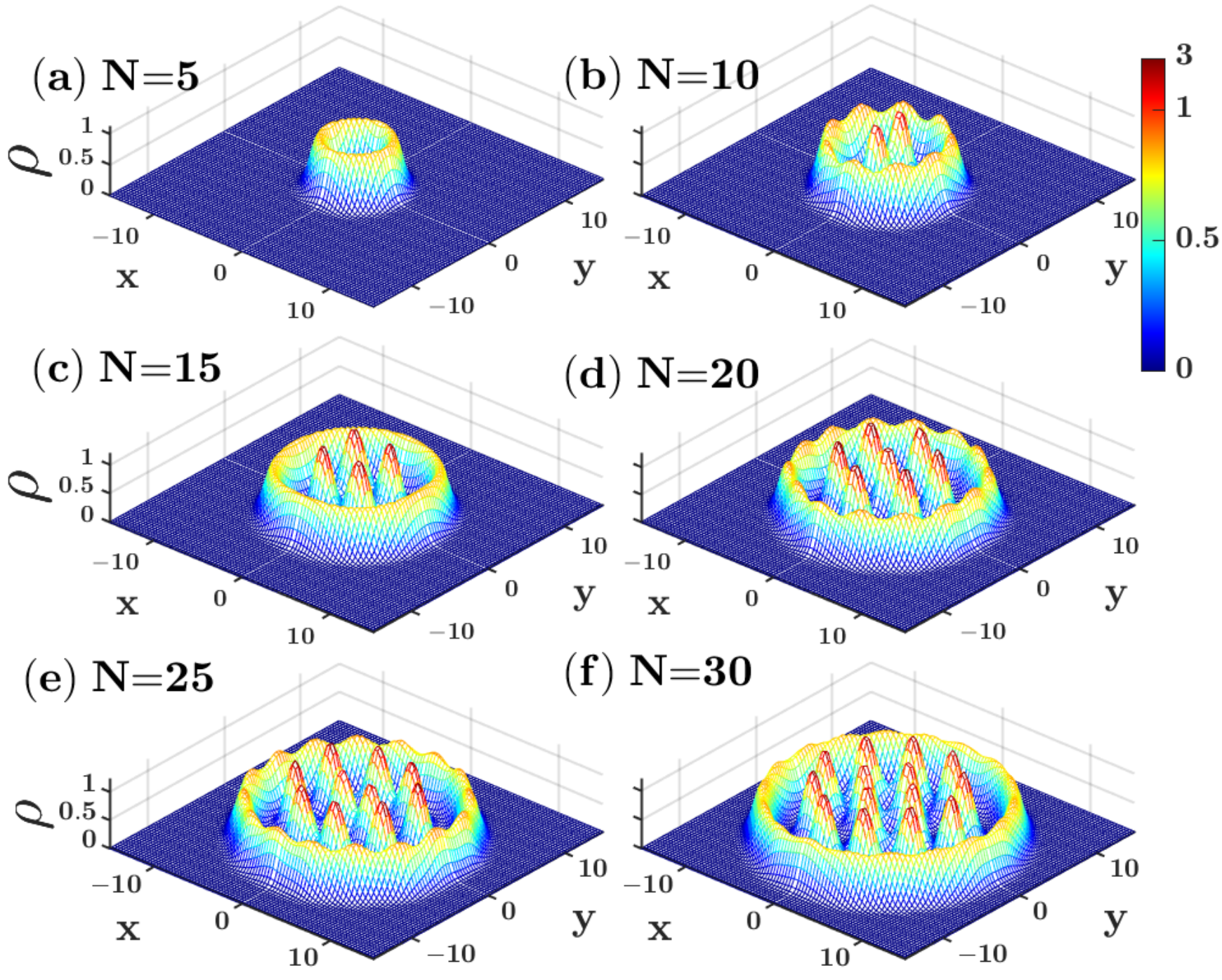}
\caption{The ground state density $\rho$ for different particle numbers $N=5, 10, 15, 20, 25, 30$ with $\varsigma=1.5$ and a uniform background charge of $\nu_{\mathrm b}=1/3$. The qualitative features of a crystalline structure in the bulk and a liquid-like ring along the edge are consistent with those in the large system of $N=40$ shown in Fig.~\ref{XCcompare}(c). We have used $L_x=L_y=30\; l_B$, $N_x=N_y=180$, $k_B\tau=0.01\frac{e^2}{\epsilon l_B}$. Other parameters are the same as those in Fig.~\ref{XCcompare}.}\label{Ncompare}
\end{figure}

We have also investigated the behavior as a function of the system size. Away from the transition region, we find that the nature of the ground state is not sensitive to system size. This is illustrated in Fig.~\ref{Ncompare}. Here, the qualitative features of the solution in Fig.~\ref{XCcompare}(c) are retained in smaller systems. For the systems with the same number of particles in Fig.~\ref{Ncompare}, the solutions remain liquid-like for $\varsigma=1$ and crystal-like for $\varsigma=2$ (results not shown).

\begin{figure*}[t]
\includegraphics[width=6.5in]{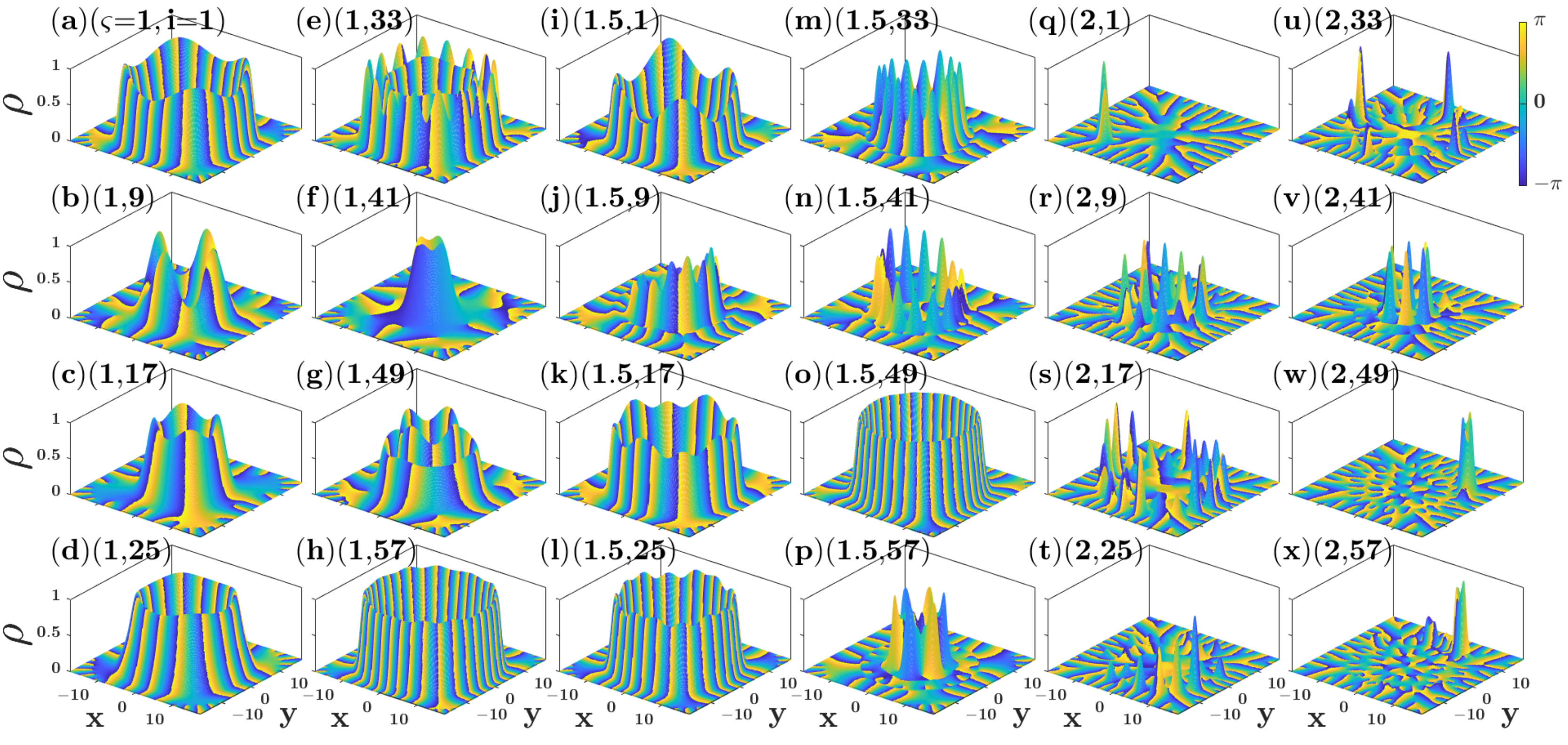}
\caption{
This figure shows the wave functions of single-particle Kohn-Sham (KS) orbitals for $\varsigma=1, 1.5, 2$. For each value of $\varsigma$, we show eight orbitals of increasing energy, with state indices $i=1 ,9 \ldots, 57$.
The numbers in the parenthesis represent the pair $(\varsigma, i)$. The height and color in the plot represent the magnitude and phase the KS orbitals, respectively. The broken rotational symmetry in the magnitude of a wave function indicates that angular momentum $L_z$ is no longer a good quantum number. In the liquid-like phase in (a-h), the expectation value of $L_z$ (which can be estimated from the number of phase windings over the azimuthal angle) is positively correlated with the average radius of an orbital. This correlation is absent in the crystal-like phases, where KS orbitals are delocalized over a few crystalline peaks. We choose the background charge at $\nu_b=1/3$, $N=40$ and $k_B\tau=0.01 \frac{e^2}{\epsilon l_B}$. Other parameters are the same as those in Fig.~\ref{XCcompare}. }\label{single}
\end{figure*}

\section{Conclusions}\label{summary}

We have studied how the strength of the XC potential between composite fermions dictates their state. For this purpose we develop a numerical procedure to solve the KS equations of CFs in a fashion that allows for crystalline solutions. Our primary finding is that the state evolves from a liquid-like state to a crystal-like state as the strength of XC energy increases.

We mention again that our study is not to be taken as a quantitative treatment of the physics of crystals in the FQHE regime. A notable limitation is that we only consider states of composite fermions carrying two vortices, and do not consider the possibility of a crystal or a liquid of electrons, or of composite fermions with greater number of attached vortices (as might be relevant in regions of small densities). 

An obvious direction for future study will be to build better XC potentials that apply to a larger range of filling factors and also include the effects of finite thickness and Landau level mixing. It is possible that the strength of the XC energy of composite fermions relative to their cyclotron energy may also depend on the filling factor, which may be relevant to the formation of a crystal at low fillings. It would be interesting to apply the DFT method to study the edge structure, the effect of disorder and/or anisotropy, spin physics, screening, and of fractional quantum Hall effect in mesoscopic devices.

\section*{Acknowledgements}
Y. H. thanks Junyi Zhang and Jiabin Yu for helpful discussions. J.K.J. thanks Steve Kivelson for an insightful discussion. The work at Penn State was made possible by financial support from the US Department of Energy under Grant No.~DE-SC0005042. Y. H. acknowledges partial financial support from China Scholarship Council. Computations for this research were performed on the Pennsylvania State University’s Institute for Computational and Data Sciences’ Roar supercomputer.

\appendix

\section{Discussion of numerical stability of the results}\label{error}

In exact diagonalization studies, the ground state of FQHE on a disk geometry is assumed to be an eigenstate of angular momentum, which therefore is rotationally symmetric.
The liquid or crystal nature of the ground state can be seen by studying the structure of the pair correlation function~\cite{Jeon04b}. However, once the rotational symmetry is spontaneously broken, the system can pick one of the infinitely many possible ground state configurations that no longer preserves angular momentum as a good quantum number. These configurations are related to each other by a rotation around the origin. Ideally this is expected to be the case for our KS solutions. (We refer the readers to Ref.~\onlinecite{Perdew21} for discussions of symmetry-breaking solutions in DFT calculations.) However, our choice of the square open boundary and our lattice discretization effectively break the rotational symmetry. We discuss in this appendix various numerical tests to show that these do not affect the nature of the state in an essential manner.

We have examined the effect of the square open boundary. We have found that adding a circular potential wall of infinite height that is internally tangent to the square boundary, or expanding the size of the square boundary by a factor of two, leaves the essential features of the solutions unchanged.

Discretization reduces the rotational symmetry of space into a four-fold rotation $C_{4}$ and the mirror reflection, which are exact symmetries regardless of the lattice constant. Interestingly, these symmetries can be broken in the KS solutions, as is evident in the density plots of crystalline solutions [for example, see Fig.~\ref{kBTcompare}(c)]. Even in the liquid solutions, the $C_{4}$ symmetry is broken by the KS orbitals, as shown in Fig.~\ref{single}. In fact, the KS orbitals can form rather complex structures. In the liquid-like phase, each KS orbital retains a ring-like shape around the origin but no longer preserves the angular momentum as a good quantum number. In the crystal-like phase, the KS orbitals typically do not respect any of the symmetries. We note that the KS orbitals are delocalized over several crystalline sites, hinting at a highly correlated nature of the crystal.

We have tested that for any solution of the KS equation, the densities related by the $C_4$ or the mirror symmetries are also valid solutions. Which of the degenerate solutions is obtained depends on the initial input. Furthermore, for the crystal phase, depending on the initial conditions, we can also obtain solutions that are not related to one another by rotation, suggesting the presence of many nearly degenerate solutions in the continuum limit. Nonetheless, all of the solutions for a given set of parameters are in the same phase.

We need to make sure that our lattice is fine enough to capture the physics in the continuum limit. To reduce the numerical expense, the results above are obtained with a lattice resolution of $a_x=a_y=l_B/6$. We have tested that going to a resolution of $l_B/10$ or $l_B/15$ does not alter the results appreciably. In the crystalline phase, the number of nearly degenerate solutions increases as we go to finer lattices, but for all cases that we have studied, going to a finer lattice does not change the liquid or crystal nature of the solution.

\end{document}